\documentclass[epj]{svjour} 
\usepackage{graphics}
\usepackage{epsfig}
\usepackage{amssymb} 
\begin{document}

\title{Transients in sheared granular matter}
\author{Brian Utter \thanks{\emph{e-mail:} utter@phy.duke.edu}
\and R. P. Behringer}

\institute{Department of Physics and Center for Nonlinear  
and Complex Systems, Box 90305, Duke University, 
Durham, NC 27708  USA}
\date{\today}

\newcommand{\ea}{{\it et al.}}
				
\abstract{
As dense granular materials are sheared, a shear band and an
anisotropic force network form.  The approach to steady state
behavior depends on the history of the packing and the existing force
and contact network.  We present experiments on shearing of dense 
granular matter in a 
2D Couette geometry in which we probe
the history and evolution of shear bands by measuring particle
trajectories and stresses during transients.  We find that when
shearing is stopped and restarted in the same direction, steady state
behavior is immediately reached, in agreement with the typical
assumption that the system is quasistatic.  
Although some relaxation of the force network is observed when 
shearing is stopped, quasistatic 
behavior is maintained because the contact network remains essentially 
unchanged. 
When the direction of
shear is reversed, a transient occurs in which stresses initially
decrease, changes in the force network reach further into the bulk,  
and particles far from the wheel become more mobile.  This
occurs because the force network is fragile to changes transverse to
the force network established under previous shear; particles must
rearrange before becoming jammed again, thereby providing resistance
to shear in the reversed direction.  
The strong force network is reestablished after displacing the 
shearing surface $\approx 3d$, where $d$ is the mean grain diameter.  Steady 
state velocity profiles are reached after a shear of $\lesssim 30d$.
Particles immediately outside of
the shear band move on average less than 1 diameter before becoming
jammed again.  We also examine particle rotation during this transient
and find that mean particle spin decreases during the transient, which
is related to the fact that grains are not interlocked as strongly.
}


\maketitle

\section{Introduction}

Common features of dense, sheared granular materials include the
formation of shear bands, which is related to Reynold's dilatancy, and
the presence of directionally oriented networks of force chains.
While the steady state behavior of dense granular shear flows 
has been the subject of numerous studies, 
research on transients have typically focused on 
the effects of cyclic loading 
and rotation of applied principal stresses 
in bulk granular samples \cite{engineering-transients}.
Much less is known about the approach to steady state and the
effects of changes in imposed shear at the grain scale 
\cite{Losert.ea:02:Transient,Toiya:04:Transient,Calvetti:97:Experimental}.

It is known that forces in compact granular materials are carried
preferentially along a network of force chains in which a minority of
the grains carry a majority of the
load\cite{Howell.ea:99:Fluctuations}.  Imposing shear establishes an
oriented network with force chains aligned preferentially so as to
resist the applied
shear\cite{Howell.ea:99:Fluctuations,Veje.ea:99:Kinematics}.  
The force network is also correlated with the network of particle
contacts, i.e. texture, although 
one can be highly anisotropic without the other  being 
so \cite{Geng:03:Green}.
Small changes in the
contact network are often sufficient to induce substantial changes in
the force network; therefore, the force network can be a sensitive measure 
of changes in granular systems. 
The anisotropy in the force
network impacts the trajectories of individual particles during
shearing \cite{Utter:04:Self} 
and creates more mobile and less mobile directions based on
shearing history.

Grains are typically confined along certain
jammed directions and more free to move along other fragile directions
\cite{Cates.ea:98:Jamming}.  
A configuration is jammed if particle rearrangement does not 
occur under the applied load.  Increasing shear stress can often lead to 
unjamming and flow \cite{Liu:98:Nonlinear}.  
In addition, a configuration can be unjammed by a small load 
in a different orientation than the  
original stress.  
An analogous situation is a tall stack of blocks which can sustain 
additional vertical load but easily collapses when pushed from the 
side. This is referred to as fragility.

When shear is applied to a uniform state, there must be a transient
during which the steady state response is achieved.  If the imposed
shear direction is altered, there must be additional transients during
which the force network, texture, and mean particle velocities adjust.

The formation of an oriented network from an initially homogeneous
state can be understood by the following argument\cite{Geng:03:Green}.
Small-angle shear deformation 
through an angle $\phi$ can be decomposed into three
parts.  Two of these are compression and dilation along axes
oriented at $45^\circ$ to the direction of shearing.  The third is a
rotation by $\phi/2$.  Force chains form primarily along the
compressional direction.  In the case of continuous shear deformation,
such as that achieved in a Couette apparatus, the heuristic description of the
force network also follows from this argument. In this case, the force chains
form in response 
to the initial shear at roughly $45^\circ$ to the shear direction, 
with an orientation that tends to resist further shearing.  However, under sufficiently 
strong shear, the network cannot resist the motion of the shearing surface; 
it fails as particles roll and
slip.  Still, on average, the orientation of the strong network remains.

In this paper, we present experimental results on transient behavior
in a 2D Couette shearing experiment.  The grains are photoelastic
disks which allow measurement of the forces at the grain scale,
thus revealing the force chain
structures.  In these experiments, we study the transients after
changing the imposed shear.  We generate these transients in one of
four ways: 
1) by imposing shear on a homogeneous packing; 
2) by stopping steady shear and observing the relaxation to a static packing;
3) by stopping steady shear, then restarting in the same direction;
or 4) by reversing
the direction of shear after steady state behavior is reached.  This
work provides a 2D analogue to the work by Losert \ea
\cite{Losert.ea:02:Transient,Toiya:04:Transient}.

If the system is stopped and shear is re-applied in the same direction, 
steady state behavior should more or less immediately be
produced (assuming slow, quasistatic shearing).  
That is, aside from some small
effect from aging of the force network, the texture should provide a
memory of where the system was stopped.  If the system is sheared in
the opposite direction after establishing a particular texture, a
transient should be observed, as the force network is reestablished
so as to once again resist shear, after which the particles are again 
close to a jammed state.

Although one expects force chains to be strong along their primary
direction, they are fragile in the transverse
direction\cite{Radjai.ea:98:Bimodal,Cates.ea:98:Jamming}
as noted above. The
experiments by Losert \ea
\cite{Losert.ea:02:Transient,Toiya:04:Transient}, using
particle tracking techniques, underscore these expectations through
their studies of transients in a 3D Couette flow.  In their
experiments, restarting shear in the same direction quickly reproduced
steady state behavior, while reversing the shear direction created a
transient in which particles far from the shearing wheel moved faster
than at steady state.  They attributed this increased mobility to a
softness of granular matter in the direction perpendicular to the
principle force axis in agreement with the picture of jammed and
fragile directions.  
Similar results by Calvetti \ea were 
attributed to an anisotropic 
contact network induced by the initial shear \cite{Calvetti:97:Experimental}.
Although both of these interpretations are quite reasonable,
these experiments 
could not directly observe changes
in the force network. 
The experiments described here provide imaging
and quantitative data for the forces at the grain scale. 

\section{Experimental Details}

\begin{figure}
\includegraphics[width=3.3in]{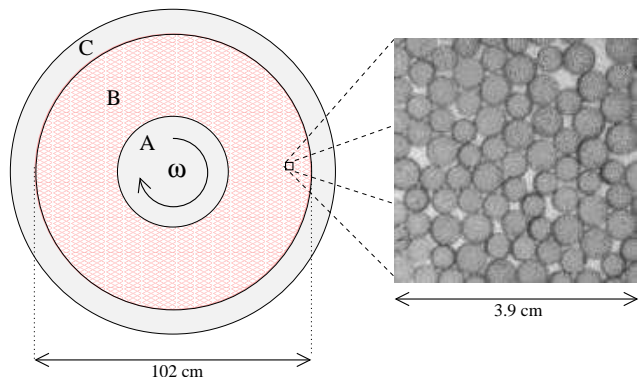}%
\caption{\label{expt-schematic} Top view of the experimental
apparatus.  The shearing wheel (A) rotates at angular frequency
$\omega = 2\pi f$ shearing the granular material (B).  The grains are
contained by a stationary outer ring (C).  On the right, an expanded
view of the grains is shown.  }
\end{figure}

We show a schematic of the experimental setup in
Fig.~\ref{expt-schematic}.  The granular
material (B) is a bidisperse mixture of photoelastic disks (thickness
0.32 cm) lying flat on a 2D Plexiglas surface.  There are
approximately 40,000 grains in a ratio of three small (diameter $d_S$
= 0.42 cm) to one large ($d_L$ = 0.50 cm) that are contained by an
inner shearing wheel (inner radius, $r_i \approx 20.5$ cm) (A) and a
stationary outer ring (outer radius, $r_o \approx 51$ cm) (C).  The
shearing wheel is rough at the particle scale and rotates at a
frequency $f$ of 0.1-10 mHz or a speed $v$ $\approx 0.013-1.3$ cm/s at
the shearing surface.  
The data presented here is for $f$ = 1 mHz and 
a mean packing fraction of $\gamma \approx$ 0.76
which can be changed between experiments by adjusting the number of
particles.  We image a subsection of the apparatus ($\leq$ 10\%) over
long times (see images below).  Each particle has a line across its
diameter to easily measure its position, orientation, and size.
Friction between the grains and the surface on which they lie is
decreased by applying a fine powder and has been measured to be two
orders of magnitude smaller than the force of a typical force chain
\cite{Howell.ea:99:Stress}.

We measure stresses by a ``gradient squared method''
\cite{Howell.ea:99:Stress} in which we calculate the mean square gradient
$\langle{G^2}\rangle$ of the image intensity at each pixel
and average over the region of interest.  
This technique is based on the properties of
photoelasticity in which increased stress on a particle illuminated
between crossed right and left circular polarizers leads to higher
density of bright and dark bands and a larger spatial gradient in
transmitted light intensity near a contact.  We have previously
established that $\langle{G^2}\rangle$ is proportional to the mean force
on a particle for the range of forces appropriate to these experiments
\cite{Howell.ea:99:Stress,Geng:03:Green}.

The experiment is initially run for multiple revolutions of the
shearing wheel to establish a steady state profile.  The shearing is
then abrubtly stopped.  The experiment can then be restarted in the
same direction (same shear) or in the opposite direction (reverse
shear).  This procedure is repeated multiple times to observe average
behavior, since fluctuations dominate for individual runs,
particularly in the force network. We present measurements of mean
particle motion and stresses below.

\section{Experimental Results}

\subsection{Overview and Organization of Results}

Below, we describe several transient processes.  The first of these is
when shear is applied to a homogeneous packing.  The second occurs
when established steady state shearing is stopped.  We then consider
what happens if shearing is stopped, allowed to relax for a modest
length of time, and then restarted in the same direction.  Finally,
the main focus is determining what happens when steady state shearing
is stopped, and then shearing is restarted in the reverse direction.

\subsection{Shearing a Homogeneous Packing}

\begin{figure}
\begin{center}
\includegraphics[width=2.8in]{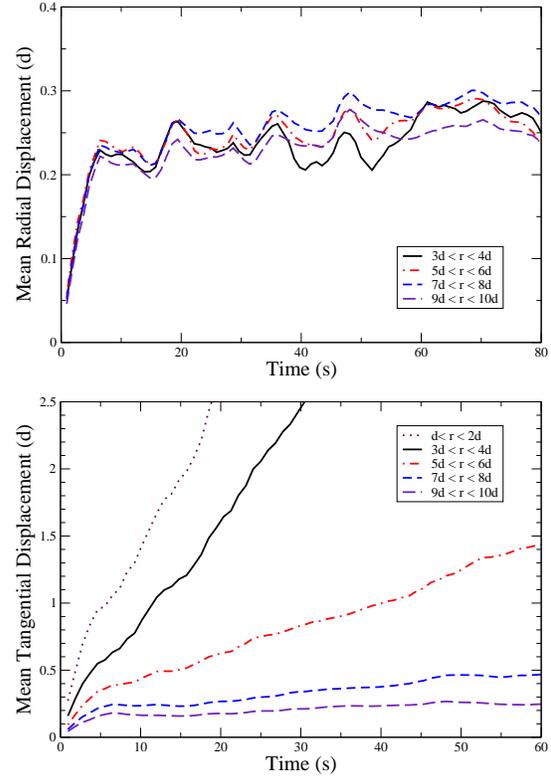}%
\caption{\label{distVSt-init} Mean radial and tangential displacements
during a transient starting from a uniform spatial distribution of
particles ($f$ = 1mHz).  After a transient of 5-10 s (shearing surface
displacement = 3 d), a steady state is achieved.}
\end{center}
\end{figure}

We first consider the case when the grains are initially uniformly
distributed before shearing is applied.  In Fig.~\ref{distVSt-init},
we show mean displacements of particles at different radial distances from
the shearing surface.  
We bin data in the radial direction and plot mean displacement 
versus time for each bin.
As expected, the disks are quickly pushed
outwards to larger values of $r$ as the granular material dilates 
at the shearing surface and the shear band forms.  
The tangential displacement also
shows a fast response before the shear band forms and particles achieve a
slower steady state average velocity.  The shear band forms very
quickly, within 5-10 s, corresponding to a displacement of under $3d$
of the shearing surface.  After this transient, the evolving
force network is qualitatively similar to that seen at steady state.

\subsection{Relaxation}

We now describe the relaxation after steady state shearing is abrubtly
stopped. Although it may be expected that relaxation would not occur,
that is not the case, as shown below.  Other recent results have
revealed a long time logarithmic relaxation of the force network
\cite{Hartley.ea:03:Logarithmic}.

\begin{figure}
\begin{center}
\includegraphics[width=2.8in]{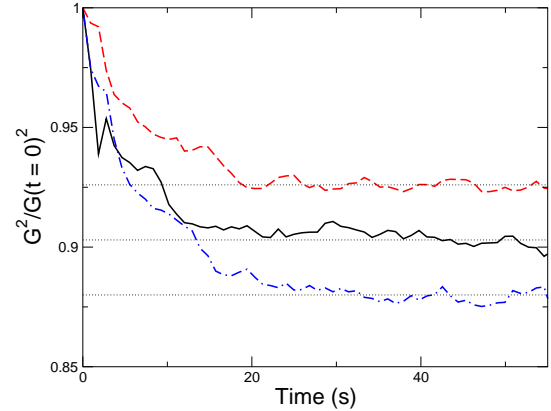}%
\caption{\label{rlx123} Decay in mean stress vs. time.  Stress is
measured using a ``gradient squared method'' as described in the text.
In three instances, we stopped the shearing (t = 0) at a point of large
stress within the field of view, and we then observed the relaxation 
of stress.  }
\end{center}
\end{figure}

\begin{figure*}
\begin{center}
\includegraphics[width=5.3in]{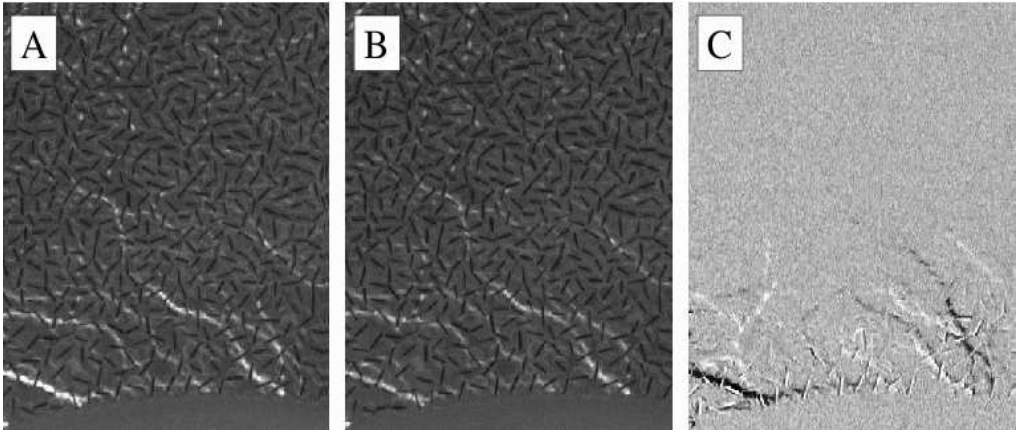}%
\caption{\label{rlx2543-593} Images shown (A) immediately after
stopping steady shear to the left and (B) 50 seconds later.  (C) The
difference between the center and left images shows that particle
motion is small and limited to grains close to the shearing wheel.
Stronger force chains have weakened (black) displacing the force to
other chains (white).  }
\end{center}
\end{figure*}

Fig.~\ref{rlx123} shows the measured stress as a function of time in
three typical cases in which steady state shear was stopped at t = 0.
In each case, we chose to stop the shearing wheel during a period of
large stress.  There was clearly a relaxation of stress over a few
10's of seconds.  During this time, the overall force network remained
largely unchanged, although a few of the stronger force chains broke
or relaxed.
We show the network relaxation explictly in Fig.~\ref{rlx2543-593}.
Here, the images correspond to when shear is stopped
(\ref{rlx2543-593}A), after 50 s (\ref{rlx2543-593}B), and
the difference between the first two images (\ref{rlx2543-593}C).
In Fig.~\ref{rlx2543-593}C, white indicates an increase of force, and dark
indicates a decrease of force.  In addition, the bars on some of the
particles close to the shearing surface shifted slightly.  The
relaxation was confined to the shear band with small displacements
observed in the grains within the first few layers of particles.  Note
that there was very little motion outside the first few layers and,
even close to the shearing surface, measureable changes in the contact
network were small.

\subsection{Restarting Shear in the Same or Reverse Direction}

In this section, we study the effect of restarting shear 
either in the same or opposite direction as a previously 
applied shear.  
In particular, we ask what transients occur before reaching steady 
state.  We 
show that restarting shear in the same direction leads to 
nearly immediate recovery of steady state behavior.  
Restarting shear in the opposite direction leads to a transient of 
approxiamtely 50-100 s (15-30 $d$ displacement of shearing surface)
during which particles outside of the shear band are more mobile and
observed stresses reach further into the bulk.

\begin{figure}
\begin{center}
\includegraphics[width=2.2in]{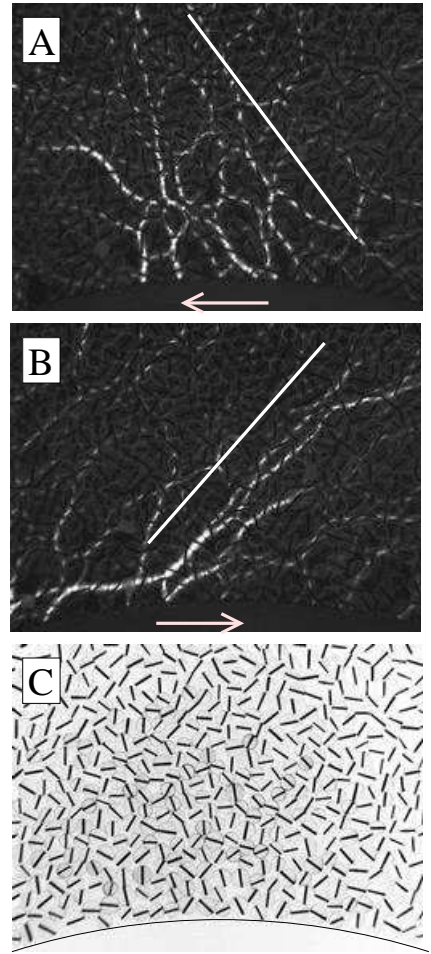}%
\caption{\label{stressimages} 
Stress images during (A) same shear and (B) reverse shear.  The arrow
indicates the shearing direction and the solid lines indicate the
approximate orientation of the strong force network.  (C) An image
showing the grains at the same magnification. }
\end{center}
\end{figure}

When shearing is reversed, the anisotropic force network must
readjust.  Fig.~\ref{stressimages} shows typical force chains during
steady state shearing along an initial direction (\ref{stressimages}A) 
and the force network with the shear reversed
(\ref{stressimages}B).  The white arrow represents the shear
direction and the straight lines indicate the approximate orientation
of the large force chains.  As described above, these form at
approximately $45^o$ to the direction of shear, so as to resist the
shear force.  The fragile directions are then at $90^\circ$ to these
lines \cite{Radjai.ea:98:Bimodal,Cates.ea:98:Jamming}.  Note that upon
reversing shear, the strong force direction switches to approximately
align with the previous fragile direction.  For comparision,
Fig.~\ref{stressimages}C shows an image of the particles at the
same magnification as that of the photoelastic force images.

\begin{figure}
\begin{center}
\includegraphics[width=2.8in]{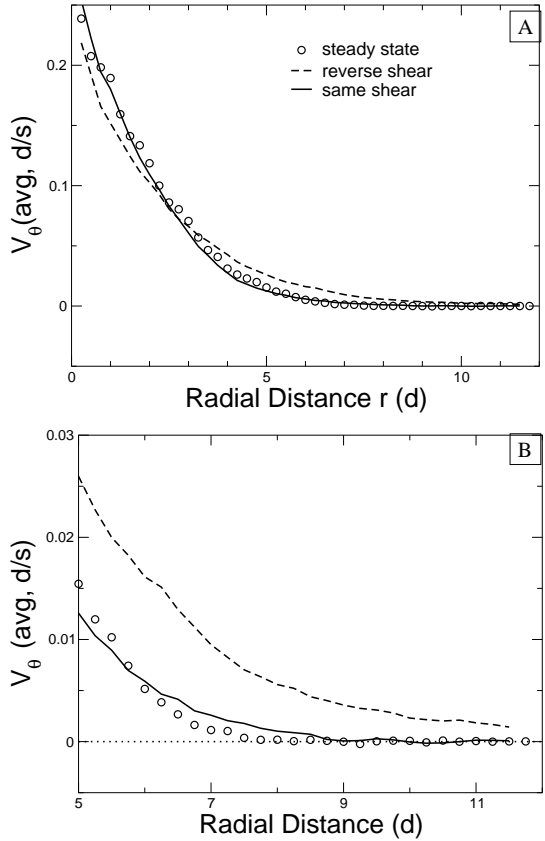}%
\caption{\label{rsroVvsR} Mean tangential velocity vs. radial distance
  from the shearing wheel for six runs ($f$ = 1mHz).  Continuing the
  same shear direction (solid line) is indistinguishable from the
  steady state profile ($\circ$) while reversing shear (dashed line)
  is clearly different.  The figure on the right is an expanded view
  of that on the left.  Particles that were jammed at steady state
  (far from shearing wheel) are now mobile, i.e. granular matter is
  fragile when forces are applied in a direction that is normal to the
  established texture.  Here, the shearing rate is $f$ = 1 mHz.  These
  data were obtained in both cases during the first 28 seconds after
  (re)imposing shear.  }
\end{center}
\end{figure}

We show velocity profiles immediately after restarting with the same
or reverse shear in Fig.~\ref{rsroVvsR}. The mean tangential velocity
$v_\theta$ is plotted versus distance from the shearing wheel in units
of the mean particle scale $d = (d_S + d_L)/2$
\cite{mean-d-note}.  
Quantities are
measured as functions of the radial distance $r$ (where $r = 0$
corresponds to the shearing surface) and averaged over azimuthal
angle, $\theta$.  Data is collected for the first 28 seconds after
restarting shear, corresponding to a displacement of $\approx 10d$ of
the shearing surface.  The curve for steady state behavior corresponds
to a longer data run (1000 s), well after the initial transient has
occurred.  The velocity profile when restarting shear in the same
direction is nearly indistinguishable from steady state behavior.
This is expected for quasistatic flows.  Although we observe some weak relaxation
of the stress network (Fig.~\ref{rlx123}) substantial
relaxation of the contact network does not occur.  Hence, the initial
velocity profiles and the profiles after restarting in the forward
direction are virtually identical.  That is, 
the force
network experiences some aging once shearing is stopped
\cite{Hartley.ea:03:Logarithmic}, but there is essentially no change in
the contact network, so the system can be
regarded as quasi-static under shearing.

When shearing is initiated in the reverse direction, there is a transient 
in the velocity profile.  Grains outside of the original
shear band become unjammed and, consequently, move more because they
are now being pushed along the fragile direction.  The effect is 
reversed for grains near 
the shearing surface.  These grains are
initially pushed away from the shearing surface because there is
initially no strong force network to resist this type of motion.
Consequently, they are not pulled along as strongly as they would 
be in an established shear band. 
Fig.~\ref{rsroVvsR}B shows an expanded view of Fig.~\ref{rsroVvsR}A.
There, it is clear that particles which were previously jammed ($7d <
r < 12.5d$) are now mobile.

\begin{figure}
\begin{center}
\includegraphics[width=2.8in]{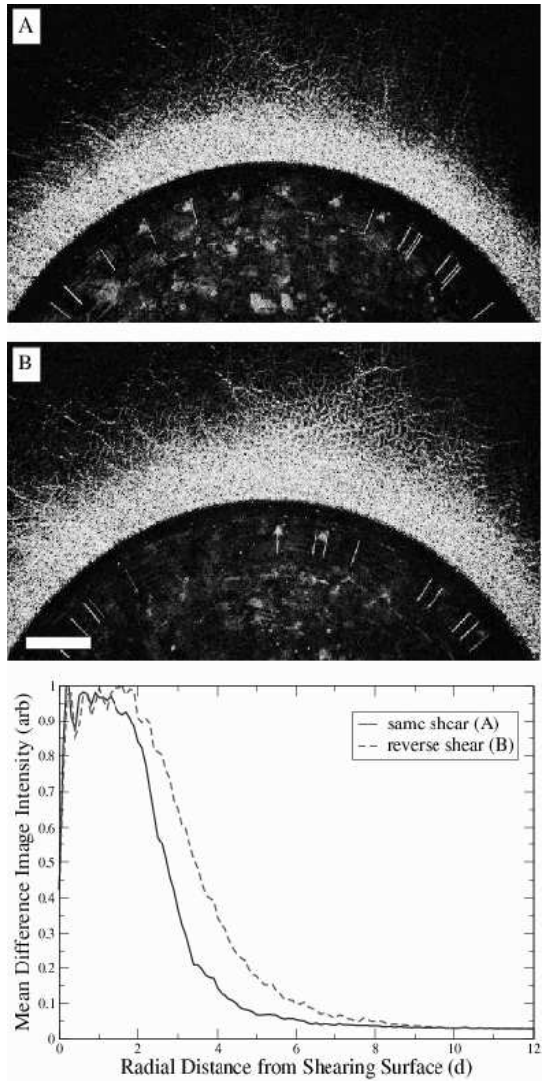}%
\caption{\label{rsro-changes30}
Difference images showing the magnitude of the changes in force network during 
transients for (A) same shear and (B) reverse shear.  Each image is 
an average of six runs.  The contrast has been enhanced to emphasize 
the differences at the edge of the shear band.  
In (C) we show the mean intensity of the original images 
versus radial distance from the shearing surface.
}
\end{center}
\end{figure}

We also examine where significant changes in the force network occur,
following the procedure used in Fig.~\ref{rlx2543-593}C.  In
Fig.~\ref{rsro-changes30}, we show difference images between the state
immediately after steady state shearing was stopped and the state 28 s
after restarting shear ($10 d$ displacement of shearing surface).  
These images show changes in the force network during the transient 
and contrast the two cases 
of shearing in the same and reverse directions.
For each run, we use the magnitude of the difference image (i.e. deviation
from grey in Fig.~\ref{rlx2543-593}C) such
that chains that form or disappear in the intervening time both
contribute equally to the resulting intensity.  
Thus, in Fig.~\ref{rsro-changes30}, white indicates a change in the force
network and black indicates where no change occured. 
In both difference images (Fig.~\ref{rsro-changes30}A and~B),
we show the average from 6 runs and adjust the contrast to 
highlight the changes 
at the edge of the shear band.  We also show a scale bar of
length $10 d$.  In each case, significant changes in the force network
are predominantly confined to the shear band ($r < 6 d$) as expected.
However, upon reversing shear, changes in the force network reach
further into the bulk, including the region where particles are
initially more mobile ($6 - 12 d$).  
In Fig.~\ref{rsro-changes30}C, we plot the (unadjusted) mean intensity versus radial 
distance from the shearing surface for the data used to construct 
Fig.~\ref{rsro-changes30}A and~B.
There it is evident that changes in the force network reach further into the 
bulk when shear is reversed. 


\begin{figure}
\begin{center}
\includegraphics[width=2.8in]{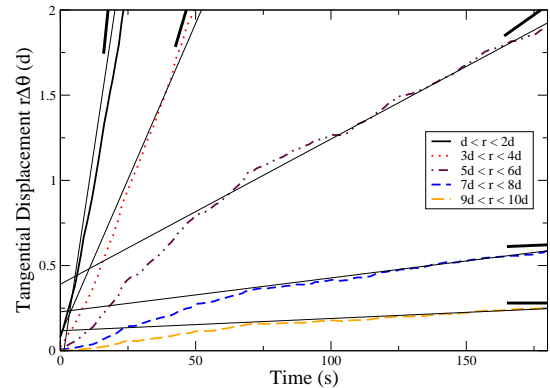}%
\caption{\label{distVSt} Mean displacement versus time for particles
at different radial distance after reversing shear.  The data is for
an average of four runs.  
Particles far from the shearing wheel ($r
\geq 7 d$) move a fraction of a particle diameter $d$ before 
reaching steady state velocity or becoming
jammed again.  
The short line segments indicate the steady state slope expected  
from Fig.~\ref{rsroVvsR}.}
\end{center}
\end{figure}

We now ask how long the increased mobility state lasts, and how far particles move
before becoming jammed again.  In Fig.~\ref{distVSt}, we show the mean
displacement versus time, averaged over four runs, for particles at
different distances from the shearing surface when the shearing
direction is reversed.  After an initial transient (t $\lesssim$ 60
s), trajectories follow $r{\Delta}\theta = v_{avg} {\Delta}t$.  The
straight lines are fits to the displacement versus time data after the
transient (100-325 s).  
The short line segments at the end of each curve 
indicate the expected slope $v_{avg}$ based on the 
steady state velocity profile in Fig.~\ref{rsroVvsR}.
Deviations between the slope of the short line segments and the 
straight lines may indicate that additional rearrangement occurs
after $t$ = 325 s.  
We also note that we compare short transient data runs to a separate 
steady state run, and slight variations in packing structure and velocity 
profiles are possible.   However, we focus on the 
most pronounced transient which occurs within the first 50-100 s.
Particles that were previously jammed or moving slowly 
initially move relatively fast 
before reaching steady state conditions.  This additional
displacement due to the increased mobility during the transient ($y$
intercept of straight line fits) is small, with displacements of less
than $1d$ needed to become jammed again.  The existence of the 
additional displacement is qualitatively the same as that found 
by Losert \ea \cite{Losert.ea:02:Transient,Toiya:04:Transient}, but 
the additional displacements are significantly smaller in our 
experiment.  The timescale for 
readjustment is 50-100 seconds, corresponding to a displacement of the
shearing wheel of 15-30 $d$.  Note that the transient 
for reversal is significantly
longer than when applying shear to a homogeneous packing
(Fig.~\ref{distVSt-init}).  In agreement with Fig.~\ref{rsroVvsR},
particles close to the wheel, initially move slower after shear reversal than at steady
state, although the lag in displacement is small compared to the
fluctuations in the steady state. 

\begin{figure}
\begin{center}
\includegraphics[width=2.8in]{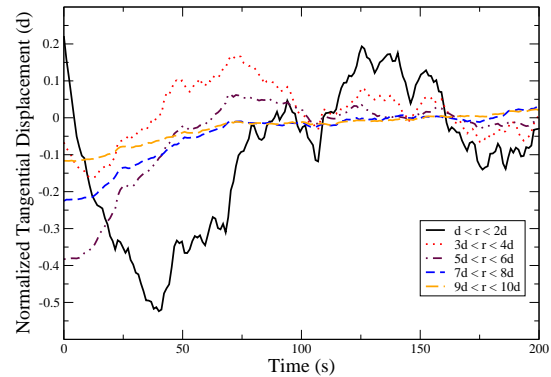}%
\caption{\label{distVSt-avg-norm} Mean tangential displacement with
the steady state fits (straight lines in Fig.~\ref{distVSt})
subtracted off.  At long times, the curves fluctuate around zero.}
\end{center}
\end{figure}

In Fig.~\ref{distVSt-avg-norm}, we have taken the data in
Fig.~\ref{distVSt} and removed the long time behavior (straight
lines).  Thus, the data in Fig.~\ref{distVSt-avg-norm} eventually
fluctuate around 0 and represent deviations from the mean steady state
over time.  At early times, negative values of the displacement show
where particles have lagged and positive values show where they have
advanced further than expected based on extrapolating 
back from the long-term steady state.
After the 50-100 s transient, not only are steady state 
velocities reached, but also
the mean motion of particles at different distances from the shearing 
is coordinated and grains at different $r$ advance or lag in unison. 

\begin{figure}
\begin{center}
\includegraphics[width=2.8in]{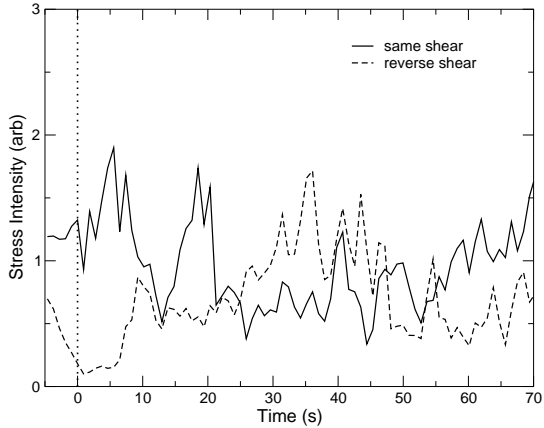}%
\caption{\label{stressVStime} 
Mean stress vs. time within the shear band ($r < 10 d$).  There is
initially no measurable stress upon reversing shear ($t < 7s$).  
Data shown is an average of six runs.
}
\end{center}
\end{figure}

We also measure the mean stress during these transients using the
$\langle{G^2}\rangle$ technique described above.
Fig.~\ref{stressVStime} shows the mean stress integrated over
all particles in $r< 10 d$, for an average of six runs.  There it is
clear that for t $<$ 7s there is no observable stress for the case
where the shear direction has been reversed.  Although the
fluctuations are different for each run, the initial quiescent period
is observed in each case and for multiple packing fractions ($\gamma = 0.755,
0.76, 0.765)$.  This is similar to the 3D results of Toiya \ea
\cite{Toiya:04:Transient}, despite the fact that in 3D, there is a
compaction of the system \cite{Losert.ea:02:Transient}.  Note that the
stress readjusts on a smaller time scale than that for velocity
relaxation.  Qualitatively, the 
force network appears similar to the steady state network after $t \simeq 10\,{\rm s}$.
Additional small adjustments of the force network might
occur during the velocity relaxation, but they are not evident due to
the strong fluctuations.

\begin{figure}
\begin{center}
\includegraphics[width=3.0in]{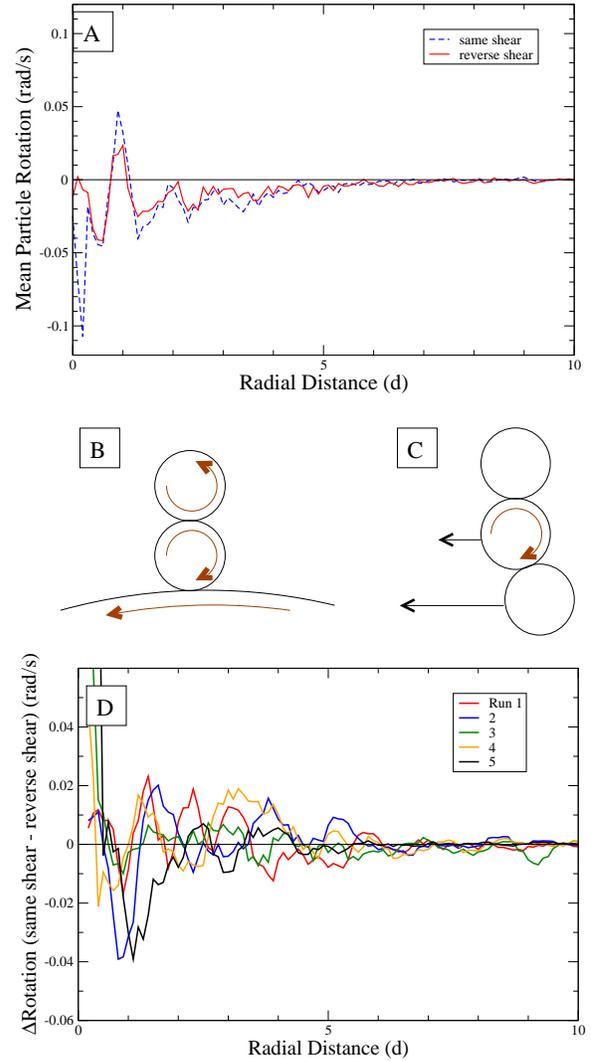}%
\caption{\label{spins} 
(A) Mean particle rotation versus radial distance. 
The mean behavior is 
influenced by (B) particles counter-rotating close
to the shearing wheel and (C) particles rotating clockwise due to shear 
($r>1$). 
In (D), we show the difference $\Delta$ Rotation which is the data for
the reverse shear rotation subtracted from the same shear rotation for 
five individual runs.
}
\end{center}
\end{figure}

We study particle rotations during the transients in
Fig.~\ref{spins}A, where we show the mean particle rotation rate versus
distance from the shearing wheel.  
The negative values close to the
shearing wheel reflect particles in contact with the wheel which are
counter-rotating (Fig.~\ref{spins}B).  If all particles in the
innermost layer were
stationary and counter-rotating with the shearing wheel, the mean
rotation rate or spin, $S$, would be $\approx -0.6$ rad/s, or in the
scaled form $S2\pi f r_i/d = -1$.  The positive peak for $r \approx 1$
is due to the counter-rotation of the second layer (Fig.~\ref{spins}B)
\cite{Veje.ea:99:Kinematics}. 
Further into the bulk ($r > d$), the mean rotation is clockwise, which 
might be expected 
for frictional particles in a shear flow (Fig.~\ref{spins}C).
That is, for the intermittent motion of a 
grain in the bulk, it is most likely that the neighbor 
further from the shearing surface is nearly stationary while the neighbor 
closer to the shearing surface moves to the left, leading to clockwise
rotation of the grain. 
The data is qualitatively similar when restarting shear in the same or
reverse directions, but the magnitudes of the peaks is somewhat
smaller during reverse shear since they are not interlocked as
strongly.  
In Fig.~\ref{spins}D, we show 
the difference in mean spin vs. $r$ (same shear - reverse shear) for 
five individual runs.  
Although the data is noisy, there is a clear trend of positive values
at $r < 0.5$ and negative values at $r \approx 1$ indicating 
the decrease in the magnitude of the 
particle spin under reverse shear seen in 
Fig.~\ref{spins}A.

\section{Conclusions}

We have studied the behavior of transients in a 2D granular Couette
shear experiment.  We find that when we restart shear in the same
direction as previous steady state shearing, no significant transients
occur.  Even though some slow relaxation of the force network 
occurs when the system is stopped, 
there is a minimal transient on restarting  
because there are no significant changes in grain positions.  
In this sense, the process is quasistatic.
When we reverse the shearing direction, transients do occur as the
force network reforms.  During this process, grains outside of the
shear band are more mobile since they are pushed along an initially
fragile direction.  Photoelastic stress images support this picture.
Mean grain displacements of less than a particle diameter are
sufficient for particles far from the shearing surface to become
jammed again and reach a nearly steady state stress structure.
Immediately after reversal, there is no measurable stress, since there
is no pre-existing force network to resist shearing.  
During the transient however, changes in the force network for reverse
shear reach further into the bulk as compared to 
what occurs for restarting shear in the same
direction.  
The strong force network is reestablished after 
displacing
the shearing surface by $\sim 3 d$.
Steady state velocity profiles are reached after a displacement of 
$\leq 30 d$.  
Reversing the shear decreases the magnitudes of
the particle spins slightly since grains are not interlocked as
strongly, but the spin profile still maintains the same qualitative
shape, with an overall reversal of sign.

{\bf Acknowledgement} We appreciate a number of helpful
discussions with Prof. Wolfgang Losert.  This work has been supported by
the National Science Foundation through grants DMR-0137119,
DMS-0204677 and DMS-0244492, and by NASA grant NAG3-2372.


\begin{thebibliography}{}
\bibitem{mean-d-note}
We use the mean of the two disk diameters $(d_S + d_L)/2$ as the relevant
  length scale. Also acceptable would be to use the mean particle size, i.e.
  $(3d_S + d_L)/4$ since our medium is a mixture of small and large particles
  in a ratio of $3:1$. In our experiment, this would decrease the value of $d$
  and rescale $v$ and $r$ by 4.3 \%.
\end{thebibliography}
\end{document}